\documentclass[a4paper,aps,prl,floatfix,twocolumn,showpacs,nofootinbib]{revtex4-1}

\usepackage[pdftex]{graphicx}
\usepackage{latexsym,amsmath}
\usepackage[english]{babel}
\usepackage[pdftex]{color}
\usepackage{hyperref}
\usepackage{microtype}

\newcommand{\upc}{Departament de F\'isica i Enginyeria Nuclear,
  Universitat Polit\`ecnica de Catalunya, Campus Nord B4, 08034
  Barcelona, Spain}

\newcommand{\av}[1]{\langle #1 \rangle}

\newcommand{\FigPath}{.}

\begin{document}

\title{Burstiness and aging in social temporal networks}

\author{Antoine Moinet}

\author{Michele Starnini}

\author{Romualdo Pastor-Satorras}

\affiliation{\upc}

\begin{abstract}
  The presence of burstiness in temporal social networks, revealed by a
  power law form of the waiting time distribution of consecutive
  interactions, is expected to produce aging effects in the
  corresponding time-integrated network.  Here we propose an
  analytically tractable model, in which interactions among the agents
  are ruled by a renewal process, and that is able to reproduce this
  aging behavior.  We develop an analytic solution for the topological
  properties of the integrated network produced by the model, finding
  that the time translation invariance of the degree distribution is
  broken.  We validate our predictions against numerical simulations,
  and we check for the presence of aging effects in a empirical temporal
  network, ruled by bursty social interactions.
\end{abstract}


\maketitle

Our understanding of the structure and properties of social interactions
has experienced a boost in recent years due to the new availability a
large amounts of digital empirical data
\cite{lazer2009life,pentland2014social}.  This endeavor has found the
necessary theoretical grounding in the newly established science of
networks \cite{barabasi2005taming}.  A first round of network studies
focused on a \textit{static} network representation
\cite{wass94,Jackson2010}, in which nodes (standing for individuals) and
edges (indicating social interactions) are constant and never change in
time.  From such static representation, a wealth of complex topological
information was obtained, concerning e.g. the presence of scale-free,
power-law degree distributions $P(k) \sim k^{-\gamma}$, large
clustering, positive degree correlations, or a distinct community
structure \cite{Newman2010}.  More recently, this framework has been
challenged by the empirical observation of a temporal dimension in
social (and other) networks, arising from the fact that social
relationships are continuously created and terminated. From these
\textit{temporal} networks \cite{Holme:2011fk}, a static representation
is obtained by means of a temporal integration of the instantaneous
interactions over a time window of width $t$, and its associated
topological properties, such as the degree distribution, $P_t(k)$, are
thus to be understood to depend on the integration time $t$
\cite{ribeiro_quantifying_2013}.  The empirical study of the temporal
aspects of social networks has unveiled an additional level of
complexity, embodied in many statistical properties showing heavy-tailed
distributions.  Remarkable among them are the distribution $\psi(\tau)$
of \emph{interevent} or \emph{waiting} times between two consecutive
social interactions, revealing the bursty nature of human dynamics, or
the distribution $F(a)$ of social \textit{activity}, measuring the
probability per unit time of establishing a new social relation, both
approximately obeying power-laws decays of the form
$\psi(\tau) \sim \tau^{-(1+\alpha)}$ and $F(a) \sim a^{-\delta}$,
respectively
\cite{Oliveira:2005fk,Barabasi:2005uq,Gonzalez:2008fk,10.1371/journal.pone.0011596,2012arXiv1203.5351P}.

This twofold nature of social interactions naturally arises the issue of
the relation between the temporal correlation properties of time-varying
networks and the topological features of their static representations.
Among others, Song \textit{at al.}~\cite{Song:2012fk} proposed an
empirical scaling theory bridging the exponents of human dynamics
patterns and social network architecture, while Perra \textit{et al.}
\cite{2012arXiv1203.5351P} considered an \textit{activity driven} model,
built upon the idea of a \textit{constant} social activity, defined as
the probability per unit time that an agent becomes active and starts a
social interaction.  The activity driven model allows to show that the
degree distribution $P_t(k)$ of an integrated network is functionally
related to the distribution of social activity by
$P_t(k) \sim t^{-1} F\left( \frac{k}{t} - \av{a} \right)$
\cite{starnini_topological_2013}.  Following this direction, in this
paper we focus on a different property of social temporal networks,
naturally expected in systems in which the addition of connections is
ruled by a non-Poissonian, power-law distribution of interevent times:
The presence of \textit{aging} behavior
\cite{Henkel2,klafter_first_2011}, which in this context translates into
a breaking of time translation invariance manifested in the dependence
of the topological properties of the integrated network on the
\textit{aging time} $t_a$ at which the integration starts.  In order to
study this possibility from an analytic point of view, we propose and
analyze a non-Poissonian activity driven (NoPad) model, in which the
waiting time between consecutive agent activations follows an arbitrary
form $\psi_i(\tau, c_i)$, $c$ being a parameter quantifying the
(possible) heterogeneity of waiting times in the population.  We compute
the topological properties of the ensuing integrated networks by
applying the hidden variables formalism
\cite{PhysRevE.68.036112,starnini_topological_2013}.  We find that if
$\psi(\tau, c)$ is a power law distribution with exponent $1+\alpha$,
then the degree distribution exponent $\gamma$ is simply related to
$\alpha$, mediated by the heterogeneity distribution $\eta(c)$. In this
model, effects of aging are clearly evident. We observe in particular
that both the degree distribution $P_{t_a, t}(k)$ and the average degree
$\av{k}_{t_a, t}$, computed in a time window $[t_a, t_a +t]$ of width
$t$, depend explicitly on the initial integration time $t_a$. Evidence
of this sort of aging is recovered in an empirical analysis of the
temporal network defined by the scientific collaborations in the journal
\textit{Physical Review Letters}, published by the American Physical
Society \cite{apsdata}.

Previous modeling efforts have shown that the concept of memory can
induce non-Poissonian interevent time distributions in temporal networks
\cite{PhysRevE.90.042805}.  Here we propose a model joining the activity
driven framework with the empirically observed bursty nature of social
interactions, which allows for a simple mathematical treatment.  The
NoPad model is defined as follows: Each agent $i$ in a network of size
$N$ is endowed with a \textit{time-dependent} activity $a_i(t)$, which
represents the probability per unit time that agent $i$ becomes active
for the first time after a time $t$ from its last activation.  When an
agent becomes active, it generates an edge that is connected to another
agent chosen uniformly at random.  Edges last for a period of time which
we assume to be infinitesimally small.  The activation of each
individual follows thus a \textit{renewal process} \cite{renewal} with a
\textit{failure} rate given by the agent's activity $a_i(t)$.  The
interevent time distribution for the node activation\footnote{We assume
  that the time between activation events is not affected by the
  reception of a connection emitted by other active agent.} is given by
\cite{renewal}
$\psi_i(\tau) = a_i(\tau) \exp \left\{ -\int_0^{\tau} a_i(\tau')
  d\tau'\right\}$.
For a time-independent activity $a_i$ we recover the original activity
driven model \cite{2012arXiv1203.5351P}, with a Poissonian, exponential
interevent time distribution.  An explicitly time dependent activity
rate leads to a non-Poissonian activity pattern, which can take the
power-law, bursty form found in human interactions.  Shifting from the
activity (failure rate) $a_i(t)$ to the equivalent waiting time
distribution $\psi(\tau)$, we define the NoPad model in a operational
way: Each agent in the network becomes active following a renewal
process ruled by a waiting time distribution $\psi(\tau, c_i)$, where
$c$ is a parameter gauging the heterogeneity of the activation rate of
the agents, and which we assume to be randomly distributed according to
a distribution $\eta(c)$.  Active agents connect to a randomly chosen
node by an edge that lasts an infinitesimally small time.

The topological properties of the NoPad model can be computed by
applying the general hidden variables formalism
\cite{PhysRevE.68.036112,starnini_topological_2013}. Hidden variables
network models consider a set of $N$ nodes, each one of them having
assigned a hidden variable $\vec{h}$, drawn from a probability
distribution $\rho(\vec{h})$.  For each pair of vertices $i$ and $j$,
$i\neq j$, an edge is created with connection probability
$\Pi(\vec{h}_{i},\vec{h}_{j})$.  The resulting network has degree
distribution
\begin{equation}
  P(k) = \sum_{\vec{h}} \rho(\vec{h}) g(k \vert \vec{h}),
  \label{eq:1}
\end{equation}
where the propagator $g(k \vert \vec{h})$ is the conditional probability
that a vertex with hidden variable $\vec{h}$ ends up with degree $k$,
and whose generating function
$\hat{g}(z \vert \vec{h})=\sum_{k}z^{k} g(k \vert \vec{h})$ fulfills the
equation \cite{PhysRevE.68.036112}
\begin{equation}
  \ln \, \hat{g}(z \vert \vec{h}) = N \sum_{\vec{h'}}\rho (\vec{h'}) \,
  \ln\left[ 1-(1-z)\Pi(\vec{h},\vec{h'})\right].
  \label{eq:2}
\end{equation}

The key point to map the NoPad model to a hidden-variable model resides
in computing the probability $\Pi_{t}(i,j)$ that two vertices $i$ and
$j$ become eventually linked at time $t$.  Following
Ref.~\cite{starnini_topological_2013}, this probability can be written
as
$\Pi_{t}(i,j) \equiv \Pi_t(r_i, r_j) = 1-
\left(1-\frac{1}{N}\right)^{r_{i}}\left(1-\frac{1}{N}\right)^{r_{j}}$,
where $r_i$ is the number of times node $i$ has become active up to time
$t$.  This number of activations is itself a random variable with
distribution $\chi_t(r \vert c)$, depending on the agent's heterogeneity
$c$ and time $t$ \cite{renewal}.  The mapping to a hidden variables
network is now clear: The hidden variables are the vector
$\vec{h} \to (r, c)$, with a probability distribution
$\rho(\vec{h}) \to \rho_t(r,c) \equiv \eta(c) \chi_t(r \vert c)$, and
the connection probability takes the form
$\Pi(\vec{h},\vec{h'}) \to \Pi_t(r, r') \simeq (r + r')/N$, independent
of $c$ and $c'$, in the limit $N\gg 1$ and $N\gg r \,, r'$.

Applying this mapping on Eq.~(\ref{eq:2}) in the limit $N\gg r \,, r'$,
we obtain a generating function $\hat{g}(z\, \vert r,c) $ with an exponential form.  
The resulting propagator is a Poisson distribution,
sharply peaked at its average value $r +\av{r}_t$, 
where $\av{r}_t = \sum_c \eta(c) \sum_r r \chi_{t}(r \vert c)$
is the average number of activation events at time $t$.
This leads, through Eq.~(\ref{eq:1}), to an approximate expression for 
the degree distribution of the integrated network at time $t$
\begin{equation}
  P_t(k) \simeq \sum_{c}\eta(c)
  \chi_{t}(k - \av{r}_t \vert c).
  \label{deg}
\end{equation}

The remaining element to close the calculation is the probability
$\chi_t(r \vert c)$, whose expression can be easily worked out in
Laplace space \cite{renewal}.  For the empirically relevant case of
heavy-tailed waiting time distributions, of the form
\begin{equation}
  \psi(t , c) = \alpha c\, (c\,t+1)^{-(\alpha+1)},  \;\;  0<\alpha<1 ,
  \label{eq:12}
\end{equation}
corresponding to a time dependent activity $a(t,c) = \alpha c /(1 +
ct)$, we can use the approximation developed in Ref.~\cite{godreche},
valid for large $r/ (ct)^\alpha$, namely
\begin{equation}
  \chi_{t}(r \vert c) \sim
  \frac{1}{(ct)^{\alpha}} e^{ \xi(\alpha, c t) \;
    r^{1/(1-\alpha)}},
\end{equation}
where $ \xi(\alpha, u) = -[1-\alpha][(\alpha/u)^{\alpha}\Gamma(1-\alpha)]^{1/(1-\alpha)}$.
From Eq.~(\ref{deg}), the degree distribution of the integrated network
up to time $t$ is given, in the continuous $c$ limit, by
\begin{equation*}
  P_t(k) 
  \sim \frac{(k - \av{r}_t)^{\frac{1-\alpha}{\alpha}}}{t} 
  \int du \,  \eta \left( \frac{u}{t} [k - \av{r}_t]^{\frac{1}{\alpha}} \right)
  \frac{e^{\xi(\alpha, u)}}{u^\alpha}.
\end{equation*}
As we will argue below, a reasonable form for the heterogeneity
distribution is a power-law one,
\begin{equation}
  \eta(c)=\dfrac{\beta}{c_{0}}\left(c/c_0\right)^{-(\beta+1)},
  \label{eq:10}
\end{equation}
with $\beta > \alpha$. From here, we obtain
\begin{equation}
  \av{r}_t \simeq \sum_c \eta(c) \frac{\sin(\pi\alpha) (c
    t)^\alpha}{\pi\alpha}
= \dfrac{\beta\sin(\pi\alpha)}{(\beta-\alpha)\pi\alpha}(c_{0}\,t)^{\alpha} \label{eq:11}
\end{equation}
 while the degree distribution is given, for $k \gg (c_0t)^\alpha$,
 by 
\begin{equation}
  P_t(k) \sim 
  (c_{0}\,t)^{\beta}
  \left(k-\av{r}_t\right)^{-1 - \frac{\beta}{\alpha}} \label{eq:3}.
\end{equation}
Eq.~(\ref{eq:3}) establishes the relation between the exponent of the
power-law degree distribution $P(k) \sim k^{-\gamma}$ and the exponent of
the long-tailed waiting time distribution, $\psi(t) \sim t^{-1-\alpha}$,
namely
\begin{equation}
  \gamma = 1 + \beta / \alpha.
  \label{eq:5}
\end{equation}
This relation, mediated through the exponent $\beta$, manifests the
relevance of the assumed distribution of heterogeneity $\eta(c)$. We can
motivate the form assumed by relating it with the empirical activity
measurements performed in Ref.~\cite{2012arXiv1203.5351P}. 
There, it was actually measured the \textit{average activity} of 
an individual $i$ over a time interval, $\bar{a_i}(\Delta t)$, 
defined as the ratio between the number of social acts 
performed by individual $i$ in the time interval $\Delta t$,
and the total number of social acts in the system in that interval. 
In the NoPad model, the number of social acts of an individual with
heterogeneity $c$ in an interval $\Delta t$ is determined by the
number of times he has become active in that interval,
 which from Eq.~(\ref{eq:11}) is given by
$\bar{r}_{\Delta t}(c) \sim c^\alpha (\Delta t)^\alpha$. 
Therefore, we have $\bar{a}(c) \sim c^{\alpha}$, independent of $\Delta t$. 
A simple transformation between probability distribution allows to write
$\eta(c) \sim F\left[ \bar{a}(c)\right] \frac{d \bar{a}(c)}{d c} \sim
c^{-1 -\alpha(\delta-1)}$.
From here, we recover the postulated heterogeneity distribution,
Eq.~(\ref{eq:10}), with an exponent $\beta = \alpha(\delta-1)$. Most
remarkably, for this value of $\beta$, the integrated network shows a
degree exponent $\gamma = 1 + \beta / \alpha = \delta$, i.e. we recover
the main result of the activity driven model, stating the equivalence
between degree and activity distributions \cite{2012arXiv1203.5351P}.

\begin{figure}[t]
  \includegraphics[width=7.5cm]{\FigPath/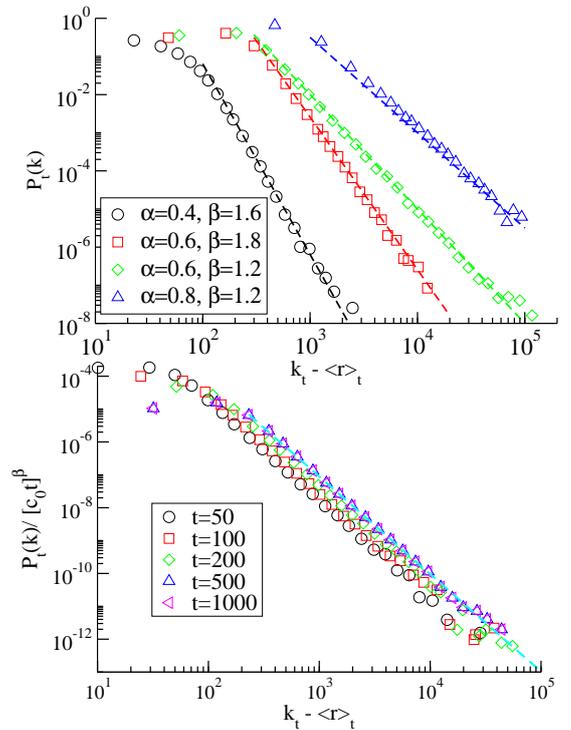}
  \caption{(color online) Top: Degree distribution $P_t(k)$ at time
    $t=10^4$ as a function of the rescaled degree $k - \av{r}_t$ from
    numerical simulation of the NoPad model with a network of size
    $N=10^6$, $c_0=1$, and different values of $\alpha$ and $\beta$.
    The exponent $\gamma$, as given by \eqref{eq:5}, is
    plotted in dashed line.  Bottom: Rescaled degree distribution
    $P_t(k) / (c_0t)^{\beta}$ for different times $t$, in networks size
    $N=10^6$, $c_0=1$, $\alpha=0.6$, and $\beta=1.2$.  The theoretical
    decay exponent $\gamma = 3$ is plotted as dashed line.}
  \label{fig:pk}
\end{figure}

In order to check our analytic predictions, we have performed numerical
simulations of the NoPad model with the waiting time and heterogeneity
distributions Eqs.~(\ref{eq:12}) and \eqref{eq:10}, respectively.  The
integrated network at time $t$ is generated as follows: To each node $i$
is assigned a heterogeneity $c_i$ extracted from the distribution
$\eta(c)$.  Then, we generate the number $r_i$ of times that each node
becomes active up to time $t$, according to the distribution
Eq.~(\ref{eq:12}).  Finally, each node $i$ is connected to $r_i$
neighbors chosen at random, avoiding multiple and self connections.  In
Fig.~\ref{fig:pk} we show the degree distribution $P_t(k)$ for different
values of the exponents $\alpha$ and $\beta$ of the waiting time and
heterogeneity distributions.  As one can see, the scaling relation of
Eq. \eqref{eq:5} is fulfilled remarkably well.  In the same Figure we
validate the scaling of the $P_t(k)$ with the integration time $t$,
Eq.~(\ref{eq:3}), showing the collapse of the degree distribution for
different $t$.

\begin{figure}[t]
  \includegraphics*[width=8cm]{\FigPath/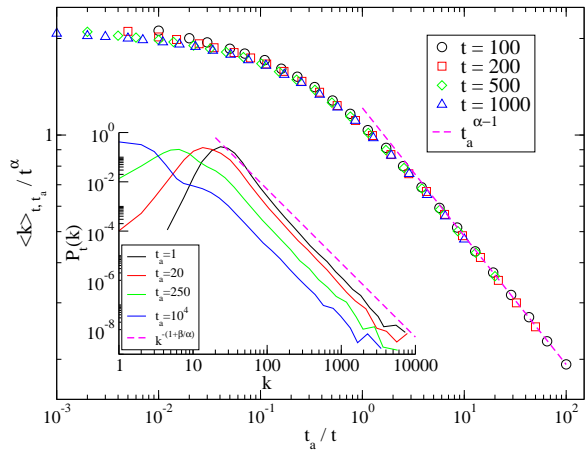}
  \caption{(color online) Rescaled average degree
    $\av{k}_{t_a, t} / t^{\alpha}$ as a function of the aging time
    $t_a$, obtained from numerical simulations of the NoPad model with a
    network of size $N=10^6$, and parameters $\alpha=0.6$, $c_0=1$,
    $\beta=1.2$, integrated for different time intervals $t$.  The
    behavior predicted for $t \ll t_a$ is shown in dashed line.  Inset:
    Degree distribution $P_{t_a,t}(k)$ of the NoPad networks of the main
    plot, for integration time $t=100$ and different aging time $t_a$.
    The behavior predicted by Eq.~\eqref{eq:5} is shown in dashed line.
  }\label{fig:aging}
\end{figure}

The dependence of the topological properties of the NoPad model on the
distribution of renewal events $\chi(r \vert c)$ readily suggests that
the model will be affected by aging effects when the waiting time
distributions have the power-law form Eq.~(\ref{eq:12}) with $\alpha<1$
\cite{klafter_first_2011}.  We check these effects in
Fig.~\ref{fig:aging} (inset), where we plot the aged degree distribution
$P_{t_a, t}(k)$ obtained from networks integrated for a time interval
$t$, started after waiting for an aging time $t_a$.  This Figure shows
that while the asymptotic shape of the $P_{t_a, t}(k)$ remains constant
for large $k$, its peak shifts to smaller values of $k$ as increasing
$t_a$. 
An analytical treatment of these aging effects is in principle possible,
using the results reported in Ref.~\cite{aging}. 
We can however easily understand them at the level of 
the aged average degree $\av{k}_{t_a,  t}$.  
Since the average degree is two times the average number of activation event, 
see Eq.~(\ref{deg}), if we consider a network integrated 
from time $t_a$ to $t_a + t$, 
we obviously have $\av{k}_{t, t_a} = 2 (\av{r}_{t_a + t}- \av{r}_{t_a}) $. 
By applying Eq.~(\ref{eq:11}) one can obtain
\begin{equation}
  \av{k}_{t, t_a} 
  \sim [(t_a  + t)^\alpha - t_a^\alpha] 
  \equiv t_a^\alpha \mathcal{F} \left( t_a t^{-1} \right).
  \label{eq:13}
\end{equation}
Thus, $ \av{k}_{t, t_a}$ exhibits a generic scaling behavior depending
on $t_a$. For $t \gg t_a$, the average degree is
independent of $t_a$, $\av{k}_{t, t_a} \sim t^\alpha \sim \av{k}_{t}$,
and aging effects are negligible.  On the other hand, for $t \ll t_a$,
the average degree decays with $t_a$ as
$\av{k}_{t, t_a} \sim t_a^{\alpha-1}$, and aging effects induce an
anomalous behavior depending on $\alpha$ \cite{klafter_first_2011}.  In
Fig.~\ref{fig:aging} we check that the Eq.~(\ref{eq:13}) correctly
reproduces the behavior of the NoPad model.

\begin{figure}[tb]
  \includegraphics*[width=7.5cm]{\FigPath/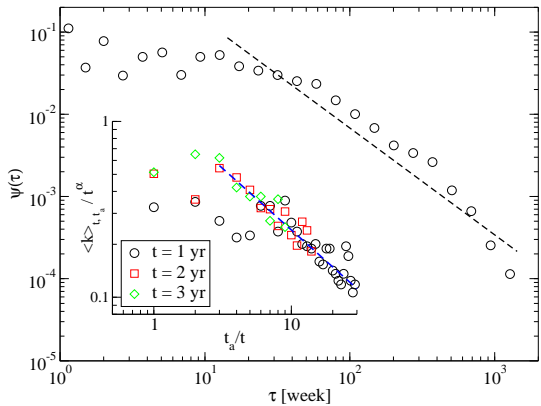}
  \caption{(color online) Waiting time distribution $\psi(\tau)$ for two
    consecutive publications in PRL by the same author, in unit of
    weeks. 
    A decay of the form given by Eq.~\eqref{eq:10}, with $\alpha = 0.3$,
    is plotted in dashed line.  Inset: Rescaled average degree
    $\av{k}_{t_a, t} / t^{\alpha}$ as a function of the aging time $t_a$
    of the PRL temporal network, integrated for different time intervals
    $t =$ 1, 2, or 3 years.  The behavior predicted by the NoPad model
    for $t \ll t_a$, $\av{k}_{t, t_a} \sim t_a^{\alpha-1}$, is shown in
    dashed line. \label{fig:prl} }
\end{figure}

We explore the possibility of aging effects in empirical temporal
networks by considering the scientific collaboration network in the
journal \textit{Physical Review Letters} (PRL), published since 1958 \citep{apsdata}.  
In this network, two authors are connected by a link if they co-authored a paper
published in PRL.  
Since PRL is weekly edited, time is measured in units of weeks.  
In order to avoid spurious effects due to effective aging of
the population considered, and single out the role of the heavy-tailed
waiting distribution, we select only those authors who published at
least one paper in any APS journal before \emph{and} after an
interval of 30 years, spanning from 1968 up to 1998.  
We then reconstruct the temporal network of the $N=677$ resulting authors, 
by considering those papers co-written by two authors in this interval,
and drawing an instantaneous edge between the authors 
at the date of the paper's publication.
In Fig.~\ref{fig:prl} we check that the waiting time distribution between
two consecutive publications of the same author has a clear heavy-tailed
form, approximately compatible with a power-law decay
$\psi (\tau) \sim \tau^{-1-\alpha}$, with exponent $\alpha \simeq 0.3$.
We then proceed to construct the integrated networks, varying the aging
time $t_a$ between 0 and 30 years and fixing the integration time as
$t = 1, 2$ and $3$ years.  In this construction it is important to
realize that the actual aging time of the network is in principle
unknown.  
Each author $i$, indeed, starts his academic life at some time $T^0_i$,
included in the observational time window between 1958 and 1968. 
Aging effects are thus observed in networks integrated over a
time window explicitly dependent on $T^0_i$ of each author considered.
This point makes extremely difficult to detect aging effects in the
degree distribution $P_{t_a, t}(k)$, also because the low activation
ratio $a_i(t)$ yields a very sparse network, with small degree values.
Nevertheless, we are able to observe aging behavior in the average
degree $\av{k}_{t_a, t} = \sum_k k P_{t_a, t}(k)$.
Fig.~\ref{fig:prl}(inset) shows the aged average degree
$\av{k}_{t_a, t}$ of the empirical data, plotted in the rescaled form
suggested by the NoPad model, Eq.~(\ref{eq:13}).  As one can see, the
data are compatible with the theoretical prediction, particularly in the
limit of large $t_a$, where we expect the actual aging time $T^0_i$ to
become small with respect to $t_a$.

To sum up, in this paper we addressed the aging effects observed in
time-integrated networks produced by bursty social interactions.
We proposed a mathematically tractable model, the NoPad model, aimed to
combine the non-Poissonian form of the waiting time distribution with
the activity-driven framework, and we developed an analytic solution for
its topological properties, through the hidden variables formalism.
Aging effects are clear in the NoPad model, as demonstrated by the
dependence of the degree distribution $ P_{t_a, t}(k)$ not only on the
integration time window $t$, but also on the aging time $t_a$ at which
we start the integration.  Inspired by the results obtained in the
model, we checked that aging behavior can also be observed in real
temporal networks.  At this respect, it is important to notice that, in
real systems, the effects purely derived by a heavy-tailed interevent
time distribution can be mixed with, and masked by, other features, such
finite-size effects, population fluctuations, s actual aging of the
individuals, memory effects, clustering or community partitioning.  The
elucidation of the contribution of all these effects in the physical
aging of temporal networks remains an open issue, deserving further
empirical and theoretical effort.

\begin{acknowledgments}
  We acknowledge financial support from the Spanish MINECO, under
  projects No. FIS2010-21781-C02-01 and FIS2013-47282-C2-2, and EC
  FET-Proactive Project MULTIPLEX (Grant No. 317532).
\end{acknowledgments}


\bibliographystyle{apsrev4-1}

%


\end{document}